\documentclass[english,aps,manuscript]{revtex4}
\usepackage[T1]{fontenc}
\usepackage[latin1]{inputenc}
\usepackage{babel}
\usepackage{graphics}

\makeatletter

\providecommand{\LyX}{L\kern-.1667em\lower.25em\hbox{Y}\kern-.125emX\@}

\newcommand{\bee}{\begin{equation}}
\newcommand{\ee}{\end{equation}}
\newcommand{\beea}{\begin{eqnarray}}
\newcommand{\eea}{\end{eqnarray}}

\makeatother
\begin{document}
\hspace*{10cm} COLO-HEP-465\\
 \hspace*{10.5cm} May 2001.

{\centering \textbf{\Large Four Flavor Finite Temperature Phase Transition
with HYP Action: Where is the First Order Phase Transition Line?}\Large \par}

{\centering \vspace{0.3cm}\par}

{\centering {\large Anna Hasenfratz\( ^{\dagger } \) and Francesco
Knechtli\( ^{\ddagger } \) }\large \par}

{\centering \vspace{0.3cm}\par}

{\centering Physics Department, University of Colorado, \\
 Boulder, CO 80309 USA\par}

{\centering \vspace{0.3cm}\par}

{\centering \textbf{Abstract}\par}

{\centering \vspace{0.3cm}\par}

{\centering We study the finite temperature phase transition of four
flavor staggered fermions with hypercubic fat link actions on \( N_{t}=4 \)
and \( N_{t}=6 \) temporal lattices. Our fat links are constructed
with hypercubic blocking (HYP) and therefore are very compact. We
present a new algorithm for simulating fermions coupled to HYP fat
links. The algorithm has a simple form based on the standard overrelaxation
and heatbath updatings for the pure gauge action. We observe that
as we increase the smoothness of the gauge fields by changing the
parameters of the blocking the very pronounced first order phase transition
of the thin link action becomes weaker and moves to physically uninteresting
values of the gauge coupling. With our smoothest HYP action we do
not find any indication of a phase transition in the accessible temperature
range on the \( N_{t}=4 \) or 6 lattices even at quark masses close
to the physical light quark mass. We argue that the observed difference
in the phase diagram is due to the improved flavor symmetry of the
smeared link actions.\par}

\vspace{0.3cm}


\vfill

\( ^{\dagger } \) {\small e-mail: anna@eotvos.colorado.edu}{\small \par}

\( ^{\ddagger } \) {\small e-mail: knechtli@pizero.colorado.edu}{\small \par}

\eject

\section{Introduction}

Understanding the finite temperature phase structure of QCD has always
been at the center of lattice studies. Recently all efforts have been
focused on two and 2+1 flavor investigations \cite{Ejiri:2000bw},
hardly any work looked at four flavor QCD \cite{Gottlieb:1987eg,Gottlieb:1989tp,Gavai:1989pr,Gavai:1990ny,Brown:1990by,Gottlieb:1991by,Engels:1997ag}.
This is not surprising, since early numerical results confirmed the
theoretical expectations of the four flavor QCD phase diagram and
computer resources were quickly directed towards the physically more
interesting two and 2+1 flavor cases. 

Based on universality Pisarski and Wilczek argued that QCD with \( n_{f}\ge 3 \)
massless quark flavors has a first order chiral phase transition \cite{Pisarski:1984ms}.
At finite quark mass the \( n_{f}^{2}-1 \) degenerate massive Goldstone
bosons can destroy the phase transition. Above some critical mass
value \( m_{max} \) one expects only a crossover. Numerical results
using thin link staggered fermions confirmed the first order phase
transition and determined the critical endpoint on \( N_{t}=4, \)
6 and 8 lattices. Since the pure gauge SU(3) theory has a first order
deconfining phase transition, at very large quark masses one might
again observe a first order transition but numerical simulations have
not explored that region. It is only a small shadow on the otherwise
satisfying picture that the numerically observed critical mass, \( m_{max} \),
does not scale as the temporal lattice size increases from four to
six. This small {}``imperfection{}'' can be explained by scaling
violations at large lattice spacing \cite{Gottlieb:1991by}. 

Just because numerical simulations seemingly agree with theoretical
expectations does not mean that we have the correct physical picture.
The Pisarski-Wilczek argument is based on the \( SU(n_{f})\times SU(n_{f}) \)
symmetry of the fermion action and the existence of \( n_{f}^{2}-1 \)
degenerate Goldstone bosons in the chirally broken phase. Staggered
fermions, on the other hand, break flavor symmetry. There is only
a remnant \( U(1) \) symmetry and a single true Goldstone boson in
staggered simulations. The other, {}``would-be{}'' Goldstone bosons
are massive pseudoscalar particles even at vanishing quark mass. The
full chiral symmetry is recovered only in the continuum limit. At
the lattice spacing corresponding to \( N_{t}=4 \) critical temperature
the flavor symmetry violation of the thin link staggered action is
substantial, the Pisarski-Wilczek scenario is not applicable. 

Chiral symmetry breaking in QCD is closely related to the topological
structure of the vacuum. Recently it was shown that the topological
susceptibility measured on two- or four-flavor thin link staggered
fermion configurations at lattice spacing \( a\simeq 0.17\,  \)fm
does not follow the theoretically predicted chiral behavior at small
quark masses \cite{Hasenfratz:2001wd}. Rather, the topological susceptibility
appears to be consistent with the quenched value independently of
the quark mass, implying that the thin link staggered fermions at
that lattice spacing do not act like two or four degenerate flavors.
The corresponding vacuum, at least regarding instantons, is closer
to the quenched than to the dynamical vacuum. One has to reduce the
lattice spacing to \( a\simeq 0.1\,  \)fm to get acceptable, though
still slightly high, values for the topological susceptibility. If
instantons are responsible for chiral symmetry breaking, the chiral
finite temperature phase transition should show similar \emph{lattice
artifacts}. The critical temperature of the chiral restoring phase
transition is expected to be \( T_{c}\simeq 150-170\,  \)MeV \cite{Engels:1997ag}.
A lattice spacing \( a\simeq 0.17\,  \)fm then corresponds to a critical
system with temporal lattice extension \( N_{t}=1/(Ta)\simeq 7-8 \),
a lattice spacing \( a\simeq 0.1\,  \)fm corresponds to a system
with \( N_{t}\simeq 12-14 \). Therefore, if the indications from
topology apply to the chiral phase transition, one has to use very
large lattices, \( N_{t}>8 \), to study the finite temperature phase
transition with thin link staggered fermions.

There is no strong dependence on the number of fermion flavors in
the above argument, assuming that the critical temperature does not
depend strongly on \( n_{f} \). Instanton model calculations indicate
that this might not be the case. Results obtained using the Interacting
Instanton Liquid Model (IILM) suggests a strong flavor dependence
\cite{Schafer:1996pz, Schafer:1998wv}. For \( n_{f}=2 \) flavors
the model predicts a second order transition at \( T_{c}\approx 150 \)MeV
in the chiral limit and only crossover at finite quark mass. With
\( n_{f}=3 \) flavors the chiral transition becomes first order and
it occurs at considerably lower temperatures, \( T_{c}\approx 100 \)MeV.
With \( n_{f}=5 \) flavors the IILM predicts that there is no spontaneously
broken phase in the massless chiral limit, that phase exists only
at non-vanishing quark mass values. The \( n_{f}=4 \) case appears
to be borderline. If a non-zero chiral condensate exists, it is very
small and occurs only at very low temperatures. These considerations
indicate that in order to reproduce the physical phase diagram on
the lattice the instanton content of the vacuum should be correctly
reproduced by the lattice action.

Flavor symmetry can be considerably improved by coupling the fermions
to fat links \cite{Orginos:1998ue,Orginos:1999cr,Knechtli:2000ku}.
Improved flavor symmetry leads to improved topology. The first results
of Ref. \cite{Hasenfratz:2001wd} indicate that fat link actions can
reproduce the quark mass dependence of the topological susceptibility
even at lattice spacing \( a\simeq 0.17\,  \)fm. It is possible that
even coarser lattices can be used with fat link fermions though it
is unlikely that a much larger lattice spacing is allowed because
instantons fall through the lattice when their radius \( \rho  \)
is comparable to the lattice spacing, \( \rho /a\simeq 1 \). Since
the average instanton radius is \( \rho \simeq 0.3\,  \)fm, configurations
with \( a\geq 0.3\,  \)fm will not have correct chiral behavior.
Because of that dynamical lattices with temporal extension \( N_{t}=4 \)
are probably too coarse to reproduce the correct continuum chiral
behavior even with chiral fermions, \( N_{t}=6-8 \) might be acceptable
with chiral or almost chiral fermions.

In a recent publication we proposed an algorithm to simulate dynamical
fat link actions \cite{Knechtli:2000ku}. In this paper we use a modified
version of that method to study the phase diagram of four flavor staggered
fermions on \( N_{t}=4 \) and \( N_{t}=6 \) temporal lattices. We
use a new type of fat link action created with a hypercubic block
(HYP) transformation \cite{Hasenfratz:2001hp}. The HYP blocking mixes
gauge links within hypercubes attached to the original link only so
that lattice artifacts due to extended smearing are minimized. The
HYP fat links are so compact that the static potential is indistinguishable
from the thin link one at distances \( r/a\geq 2 \), yet flavor symmetry
with staggered quarks is improved by about an order of magnitude.
Our results for the finite temperature phase diagram with HYP staggered
fermions deviate significantly from the thin link action predictions.
At \( N_{t}=4 \) and constant physical quark mass values the strongly
first order phase transition observed with thin links weakens and
moves deep into the strong coupling region as fattening is introduced.
It appears that the \( N_{t}=4 \) first order phase transition with
thin link staggered fermions is more a lattice artifact than a physical
phase transition. With our smoothest HYP action we do not see any
sign of a first order phase transition even at quark masses comparable
to the physical light quark mass, neither on \( N_{t}=4 \) nor on
\( N_{t}=6 \) lattices. Because of the large lattice spacing at \( N_{t}=4 \)
even a chirally symmetric action might not be reliable but the \( N_{t}=6 \)
results should be closer to the continuum behavior. We believe that
the observed difference between the HYP and thin link actions is due
to the 15 near-degenerate Goldstone bosons of the fat link HYP action
which can change the phase diagram.

There are several groups doing calculations using dynamical fat link
actions though with fat links that are less smooth than the HYP fat
links \cite{Karsch:2000ps,Bernard:2001av}. Unfortunately they did
not publish results about the finite temperature phase transition
with four flavors or on the topological susceptibility with either
two or four flavors using fat link actions. We feel that the topological
susceptibility is one of the best indicators to distinguish quenched
and dynamical configurations and verify the effect of the sea quarks
on the vacuum. Since Ref. \cite{Hasenfratz:2001wd} has only preliminary
results using a different fat link action than considered here, we
are presently investigating the topological susceptibility of the
HYP action at different lattice spacings and quark masses. Early results
show consistency with the theoretically predicted chiral behavior
at a lattice spacing \( a\simeq 0.17\,  \)fm and at a quark mass
corresponding to a Goldstone pion mass \( m_{G}r_{0}\simeq 2.0\,  \).
Detailed results will be presented in a forthcoming publication \cite{colonext}.

This article is organized as follows. In Sect. 2 we present our new
algorithm to simulate fat link actions. It is based on the standard
overrelaxation and heatbath updatings for the pure gauge action. We
describe the actions used in this study and the performance of the
algorithm. In Sect. 3 we present our finite temperature results for
four flavors of staggered fermions. We show how the phase diagram
at \( N_{t}=4 \) is changed by smoothing the gauge fields. With our
smoothest HYP action we study the phase diagram when the quark mass
is lowered and the spatial volume is increased. We also simulate the
HYP action on finer lattices at \( N_{t}=6 \). There is no sign of
a first order phase transition even with very light quarks in the
physically relevant coupling range. In Sect. 4 we conclude with a
summary.

\section{Simulations}

\subsection{Algorithm\label{ss_algo}}

Dynamical simulations of fat link actions are difficult because the
fermions couple to a complicated, extended, and, in case of projected
fat links, non-linear combination of the gauge links. In \cite{Knechtli:2000ku}
we considered fat links created by several levels of projected APE
blocking. We introduced an auxiliary gauge field for each blocking
level and coupled the fermions to the last level of auxiliary gauge
fields. 

In this article we present a new action that has no auxiliary gauge
fields and can be simulated in a simpler way. Our new fat link action
 is \begin{equation}
\label{action}
S\: =\: -\frac{\beta }{3}\sum _{p}Re\, Tr(U_{p})-tr\, \ln [M^{\dagger }(V)M(V)]\, ,
\end{equation}
 where \( U_{p} \) is the plaquette product of the thin gauge links
\( U_{i,\mu } \), \( M \) is the fermionic matrix and \( V_{i,\mu } \)
is the fat link that couples to the fermions. We denote by \( Tr \)
the trace over SU(3) color whereas \( tr \) means the trace over
space-time indices \( i \), directions \( \mu  \), spin and color.
If the fat links are constructed in a deterministic way from the thin
gauge fields, they are no longer dynamical variables and an ergodic
update of the \( U \) fields will correctly simulate the system.
Fat links can even be constructed by iterating the block transformation
several times and projecting back onto SU(3) after each blocking step.
The general algorithm to simulate the action of eq(\ref{action})
is based on a two-step decomposition of the action. First, a subset
of the thin links are updated with the standard microcanonical overrelaxation
\cite{Adler:1981sn,Petronzio:1990vx} or Cabibbo-Marinari heatbath
\cite{Cabibbo:1982zn} algorithms for the pure gauge action \( S_{G}(U)=-\frac{\beta }{3}\sum _{p}Re\, Tr(U_{p}) \).
Next, the deterministically constructed fat links are updated and
this change is accepted with probability\begin{equation}
\label{pacc}
P_{acc}(V^{\prime },V)\: =\: \min \left\{ 1,\exp [-S_{F}(V^{\prime })+S_{F}(V)]\right\} \, ,
\end{equation}
 where \( V^{\prime } \) denotes the new fat link configuration and
\( S_{F}(V)=-tr\, \ln [M^{\dagger }(V)M(V)] \) is the fermionic action.
The sequence of thin link updates must be carefully chosen to satisfy
detailed balance with respect to the pure gauge action \( S_{G} \).
The proof that this algorithm satisfies detailed balance with respect
to the full action \( S=S_{G}+S_{F} \) follows closely the proof
given in \cite{Hasenbusch:1998yb} for a similar two-step decomposition
of the action.

In the practical implementation of the algorithm we choose the subset
of thin links to be updated differently for the overrelaxation and
for the heatbath steps. For the overrelaxation we reflect all the
links within some finite block of the lattice. The location of the
block is chosen randomly but its dimensions are fixed. We choose with
probability 1/2 a given sequence of reflections within this block
or with equal probability the reversed sequence. The sequence has
to be reversed with respect to the direction and location of the thin
links and with respect to the index of the SU(2) subgroup used in
the reflection step. For the heatbath we choose the subset of thin
links randomly, i.e. we choose a random direction, a random parity,
a set of random sites and a random sequence of SU(2) subgroups. The
probability of generating a given sequence of thin link updates or
the reversed sequence is then the same, both for overrelaxation and
heatbath. From this it follows that detailed balance with respect
to the pure gauge action \( S_{G} \) is satisfied for both updatings.
We chose to update a contiguous block of links with overrelaxation
in order to propagate a change through (a portion of) the lattice.
We observed that a random sequence of overrelaxed reflections is less
efficient, especially for changing the topological charge of the configurations.
This can be understood by thinking of instantons as extended objects.
One has to change an entire region of the lattice to destroy or to
create them.

The acceptance probability eq(\ref{pacc}) contains the ratio of fermionic
determinants \( \det [M^{\dagger }(V^{\prime })M(V^{\prime })]/\det [M^{\dagger }(V)M(V)] \)
which we evaluate by the same method as described in \cite{Knechtli:2000ku}.
To make this article self-contained we summarize the steps in the
following. First, we remove the most ultraviolet part of the fermion
matrix by decomposing it as \begin{equation}
\label{redfmat}
M(V)\: =\: M_{r}(V)A(V)\quad \mbox {with}
\end{equation}
\begin{equation}
\label{amat}
A(V)\: =\: \exp \left[ \alpha _{4}D^{4}(V)+\alpha _{2}D^{2}(V)\right] \, ,
\end{equation}
where \( D \) is the kinetic part of the fermion matrix and the parameters
\( \alpha _{4} \) and \( \alpha _{2} \) are arbitrary but real.
This way we achieve that an effective gauge action\begin{equation}
\label{effgaction}
S_{eff}(V)\: =\: -2\alpha _{4}Re\, tr[D^{4}(V)]-2\alpha _{2}Re\, tr[D^{2}(V)]
\end{equation}
is removed from the determinant. The acceptance probability eq(\ref{pacc})
can be approximated by the stochastic estimator\begin{equation}
\label{stochacc}
P^{\prime }_{acc}(V^{\prime },V)\: =\: \min \left\{ 1,\exp \left\{ S_{eff}(V)-S_{eff}(V^{\prime })+\xi ^{\dagger }\left[ M_{r}^{\dagger }(V^{\prime })M_{r}(V^{\prime })-M_{r}^{\dagger }(V)M_{r}(V)\right] \xi \right\} \right\} \, ,
\end{equation}
where the vector \( \xi  \) is generated according to the probability
distribution\begin{equation}
\label{xsivec}
P(\xi )\: \propto \: \exp \left\{ -\xi ^{\dagger }M_{r}^{\dagger }(V^{\prime })M_{r}(V^{\prime })\xi \right\} \, .
\end{equation}
The parameters \( \alpha _{2} \) and \( \alpha _{4} \) can be optimized
to maximize the acceptance rate. We keep the same choice \( \alpha _{2}=-0.18 \)
and \( \alpha _{4}=-0.006 \) as in \cite{Knechtli:2000ku}.

\subsection{Actions}

In this study of the QCD thermodynamics we use four flavors of staggered
fermions and compare results obtained with three different fermionic
actions. Besides the standard action with thin link gauge connections,
we simulate two fat link actions of the form eq(\ref{action}): the
APE1 action and the HYP1 action. The simple form of the action and
of the algorithm based on overrelaxation and heatbath updatings with
respect to the gauge action allows us to simulate fat link action
with any fattening procedure. No further improvement of the gauge
and fermionic actions is used here.

In the APE1 action the fat links are constructed using only one level
of APE smearing with APE parameter \( \alpha =0.7 \) \cite{Albanese:1987ds}.
The fat link is defined as \begin{equation}
\label{ape}
V_{i,\mu }=Proj_{SU(3)}[(1-\alpha )U_{i,\mu }+\frac{\alpha }{6}\sum _{\pm \nu \neq \mu }U_{i,\nu }U_{i+\hat{\nu },\mu }U_{i+\hat{\mu },\nu }^{\dagger }]\, ,
\end{equation}
 where the projection onto SU(3) is deterministic. In the HYP1 action
the fat links are constructed using one level of hypercubic (HYP)
blocking and we refer to Ref. \cite{Hasenfratz:2001hp} for the precise
definition and the parameter choice of the HYP fattening. The HYP
fat links mix thin links within the hypercubes attached to the original
link only. The artifacts due to extended smearing, e.g. in the static
potential at short distances, are minimized with the HYP blocking.
At the same time the improvement of chiral or flavor symmetry is substantial.
In \cite{Hasenfratz:2001hp} it has also been demonstrated that one
level of HYP blocking remains effective as the lattice spacing is
decreased. Finally, the APE smearing can be considered as a particular
choice of the parameters of the HYP smearing.
\begin{table}
{\centering \begin{tabular}{|c|c|c|c|}
\hline 
Action/\( \delta _{2}(\pi ) \)&
Thin&
APE1&
HYP1\\
\hline
\hline 
\( \pi _{i,5} \)&
0.594(25)&
0.212(22)&
0.086(14)\\
\hline 
\( \pi _{i,j} \)&
0.72(6)&
0.35(5)&
0.150(24)\\
\hline
\end{tabular}\par}

\caption{The parameter \protect\( \delta _{2}\protect \) measuring flavor
symmetry violations as computed in the quenched approximation at \protect\( \beta =5.7\protect \)
for the different quark actions that we use in this study.\label{tabfsymm}}
\end{table}

\begin{figure}
{\centering \resizebox*{10cm}{!}{\includegraphics{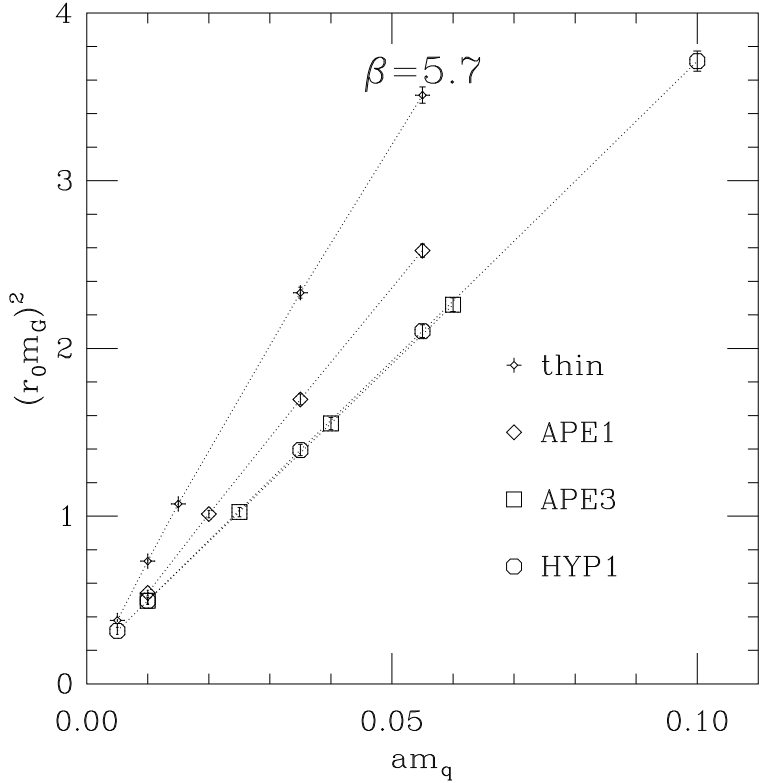}} \par}

\caption{The mass renormalization in the quenched approximation at \protect\( \beta =5.7\protect \)
for different quark actions. The APE3 action is the fat link action
of Ref. \cite{Knechtli:2000ku}. The three lowest quark mass values
with thin link action are from Ref. \cite{Gupta:1991mr}. \label{massren}}
\end{figure}

To compare the flavor symmetry violations of the standard, APE1 and
HYP1 actions we calculated the quenched pion spectrum on an ensemble
of \( \beta =5.7 \), \( 8^{3}\times 24 \) pure gauge configurations
with lattice spacing \( a\simeq 0.17\,  \)fm. As we demonstrated
in \cite{Knechtli:2000ku} the quenched result is a good indication
of the dynamical flavor symmetry violations. We use the parameter
\cite{Orginos:1998ue}\begin{equation}
\label{delta2}
\delta _{2}=\frac{m^{2}_{\pi }-m^{2}_{G}}{m^{2}_{\rho }-m^{2}_{G}}
\end{equation}
 at \( m_{\pi }/m_{\rho }=0.55 \) to quantify the difference between
the mass of the true Goldstone pion \( m_{G} \) and the mass of the
pseudoscalar particles (or pions) \( m_{\pi } \). The results are
summarized in table \ref{tabfsymm}. We compute the parameter \( \delta _{2} \)
for the two lightest non-Goldstone pions, \( \pi _{i,5} \) with flavor
structure \( \gamma _{i}\gamma _{5} \), and \( \pi _{i,j} \) with
flavor structure \( \gamma _{i}\gamma _{j} \). For a flavor symmetric
action \( \delta _{2}=0 \). The flavor symmetry violation of the
thin link action is reduced by a factor 3 with the APE1 action and
by a factor 6 with the HYP1 action at this lattice spacing. 

In figure \ref{massren} we plot the square of the Goldstone pion
mass \( m_{G} \) in units of the Sommer scale \( r_{0} \) \cite{Sommer:1994ce}
as function of the bare quark mass \( am_{q} \). In addition to the
three valence actions used in this study we plot the data for the
APE3 action (3 levels of APE smearing with \( \alpha =0.7 \) and
projection parameter \( \lambda =500 \)) of Ref. \cite{Knechtli:2000ku}.
The data for the three lowest quark masses with thin link action come
from Ref. \cite{Gupta:1991mr} and are obtained on \( \beta =5.7, \)
\( 16^{3}\times 32 \) lattices. The mass renormalization after one
level of HYP blocking is the same as after 3 levels of APE blocking.
We will use these quenched results to approximately match the physical
quark masses between dynamical simulations of four flavors of HYP
and thin link staggered fermions.

\subsection{Performance}

We would like to compare the performance of the algorithm described
in section \ref{ss_algo} for the simulation of the HYP1 action with
the standard HMC algorithm for the thin link action. We would like
to emphasize that the implementation of the algorithm for the HYP1
action is very simple and can be easily parallelized. We also optimized
our parallelized code reducing considerably the simulation time cost.

On \( 8^{3}\times 24 \) lattices we simulated the HYP1 action with
gauge coupling \( \beta =5.2 \) and quark mass \( am_{q}=0.1 \).
An approximate matching of the lattice spacing and the physical quark
mass can be achieved by simulating the thin link action with parameters
\( \beta =5.2 \) and \( am_{q}=0.06 \). In an attempt to make a
reliable comparison of the time costs, we estimated the autocorrelation
times for simple observables like the plaquette and the chiral condensate
\( <\overline{\psi }\psi > \). We found that measurements with thin
link action separated by one HMC trajectory of unit time length have
comparable autocorrelations as measurements with HYP1 action separated
by 160 overrelaxation steps and 80 heatbath steps. Each overrelaxation
step updates a block of 128 links, each heatbath step updates 200
links with an acceptance of about 20\% for both updatings. In these
physical time units the simulation of the HYP1 action is a factor
of 7 more expensive than the thin link action. 

Interestingly neither the heatbath nor the overrelaxation algorithm
looses efficiency as the quark mass is lowered. With a combination
of the two updates one can easily simulate systems at very small quark
masses that would be impractical if not impossible to simulate with
standard fermionic algorithms. We simulated on \( 8^{3}\times 24 \)
lattices the HYP1 action with parameters \( \beta =5.2 \), \( am_{q}=0.04 \)
and the thin link action with parameters \( \beta =5.2 \), \( am_{q}=0.024 \),
with approximately matched physical quark masses. The time costs are
now only a factor 3 higher for the HYP1 action as compared to the
thin link action, due to the considerable increase of conjugate gradient
steps needed in the inversion of the fermion matrix with thin links.

At a fixed lattice spacing the necessary number of overrelaxation
and heatbath updating steps between independent configurations scales
with the volume. Since each updating requires the evaluation of the
fermionic action, this gives a volume square dependence for the cost
of the algorithm. On the other hand the number of links that can be
updated with the overrelaxation or heatbath algorithms scales with
the lattice spacing, the necessary number of updating steps for a
fixed physical volume is independent of the lattice spacing. We simulated
the HYP1 action on temporal lattices with \( N_{t}=4 \) and \( N_{t}=6 \)
time extensions. The change in lattice spacing corresponds to about
a factor of 5 change in lattice volumes. At present we cannot match
the physical scales between the \( N_{t}=4 \) and \( N_{t}=6 \)
simulations. Nevertheless, comparing temperature regions where similar
changes in the observables occur (e.g. the region where the chiral
condensate starts increasing as the temperature is lowered) we notice
that we can update about 5 times as many links on the \( N_{t}=6 \)
lattices than on the \( N_{t}=4 \) lattices. This indicates that
the physical volume of the updated region remains constant as the
continuum limit is approached.

\section{Finite Temperature Results}

In this finite temperature study we performed simulations at \( N_{t}=4 \)
on \( 8^{3}\times 4 \), \( 10^{3}\times 4 \) and \( 16^{3}\times 4 \)
lattices and at \( N_{t}=6 \) on \( 16^{3}\times 6 \) lattices.
At \( N_{t}=4 \) we considered both the APE1 and HYP1 actions and
we will compare our results to thin link simulations. We used the
data from quenched spectroscopy simulations shown in figure \ref{massren}
to match approximately the quark masses of the different actions.
The bare mass values used in our simulations are listed in table \ref{mass_values}.
At the lowest quark mass value we simulated only the HYP1 action as
it became computationally very demanding to simulate both the thin
and APE1 actions. At \( N_{t}=6 \) we simulated only the HYP1 action
with a mass that corresponds to the lowest mass at \( N_{t}=4 \).
To get a feeling for the physical value of the masses, assuming a
critical temperature \( T_{c}\simeq 150-170\,  \)MeV for four flavor
QCD \cite{Engels:1997ag} and using \( Z=1 \) for the mass renormalization
factor we estimate that the lowest quark mass simulations corresponds
to \( m_{q}\simeq 6-7\,  \)MeV. Even though this estimate can easily
have a factor of two to four error, this value is still very close
to the physical light mass values.

The observables that we measure in our simulations are the average
Polyakov loop \begin{equation}
\label{polyline}
L\: =\: \frac{1}{N^{3}_{s}}\sum _{\vec{n}}Tr\prod ^{N_{t}-1}_{t=0}U_{(\vec{n},t),0}\, ,
\end{equation}
 the chiral condensate\begin{equation}
\label{psibarpsi}
\overline{\psi }\psi \: =\: \frac{1}{N_{s}^{3}N_{t}}tr\, M^{-1}(V)\: \simeq \: \frac{1}{10N_{s}^{3}N_{t}}\sum ^{10}_{i=1}R_{i}^{\dagger }M^{-1}(V)R_{i}\, 
\end{equation}
which we evaluate using 10 Gaussian random vectors \( R_{i} \) per
configuration and the disconnected chiral susceptibility \[
\chi _{\bar{\psi }\psi ,\, disc}=\frac{n_{f}}{16}N^{3}_{s}N_{t}(<(\bar{\psi }\psi )^{2}>-<\bar{\psi }\psi >^{2}).\]

\begin{table}
{\centering \begin{tabular}{|c||c|c|c||c|}
\hline 
Action/Set&
Thin &
APE1&
HYP1&
\( N_{s} \)\\
\hline
\hline 
Set 1 (\( N_{t}=4 \))&
0.06&
0.08&
0.1&
8\\
\hline 
Set 2 (\( N_{t}=4 \))&
0.024&
(0.032)&
0.04&
8, 10\\
\hline
Set 3 (\( N_{t}=4 \))&
(0.006)&
(0.008)&
0.01&
8, 10, 16\\
\hline
Set 3 (\( N_{t}=6 \))&
(0.004)&
(0.0053)&
0.0067&
16\\
\hline
\end{tabular}\par}

\caption{The bare quark mass values \protect\( am_{q}\protect \) for the
different actions that correspond to approximately the same physical
mass. An entry in parenthesis indicates that we did not simulate that
action in the data set. In the last column we list the spatial extension
\protect\( N_{s}\protect \) of the lattices in our simulations. \label{mass_values} }
\end{table}

\begin{figure}
{\centering \resizebox*{16cm}{!}{\includegraphics{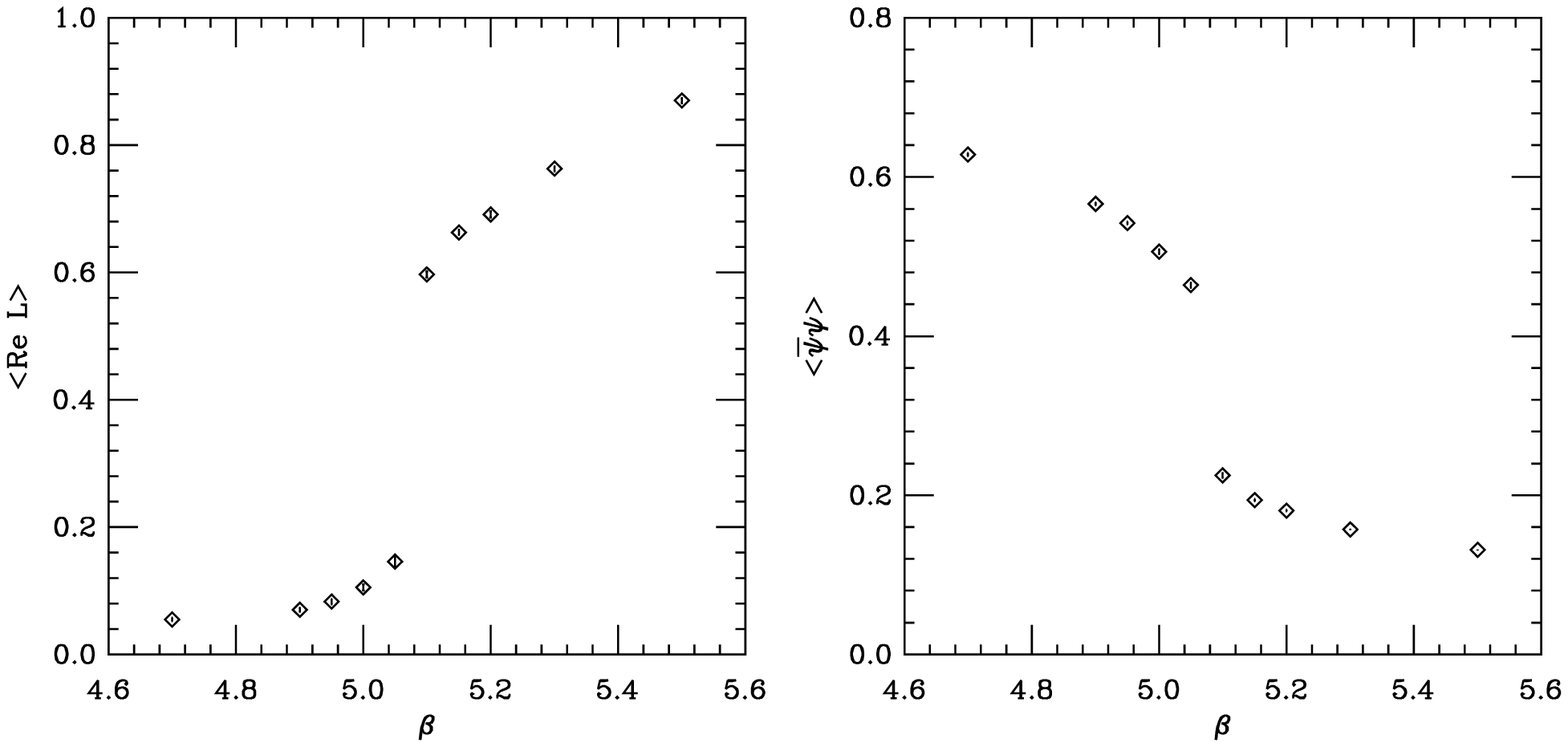}} \par}

\caption{Thin link action simulated on \protect\( 8^{3}\times 4\protect \)
lattices at quark mass \protect\( am_{q}=0.06\protect \) (Set 1 in
table \ref{mass_values}). The real part of the Polyakov line and
the chiral condensate \protect\( <\bar{\psi }\psi >\protect \) are
plotted as a function of \protect\( \beta \protect \). \label{standard}}
\end{figure}

\begin{figure}
{\centering \resizebox*{16cm}{!}{\includegraphics{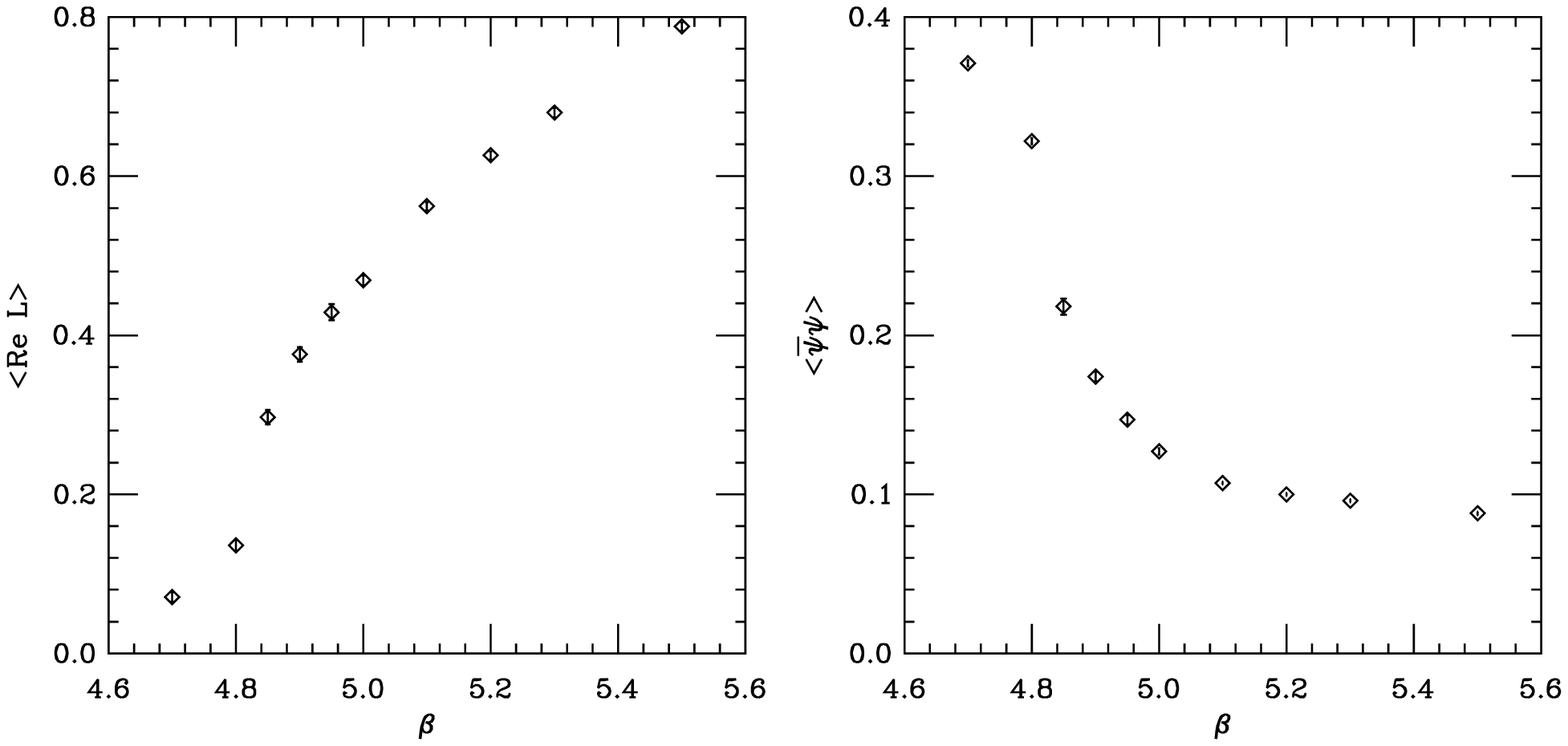}} \par}

\caption{APE1 action simulated on \protect\( 8^{3}\times 4\protect \) lattices
at quark mass \protect\( am_{q}=0.08\protect \) (Set 1 in table \ref{mass_values}).
The real part of the Polyakov line and the chiral condensate \protect\( <\bar{\psi }\psi >\protect \)
are plotted as a function of \protect\( \beta \protect \). \label{APE1}}
\end{figure}

\begin{figure}
{\centering \resizebox*{16cm}{!}{\includegraphics{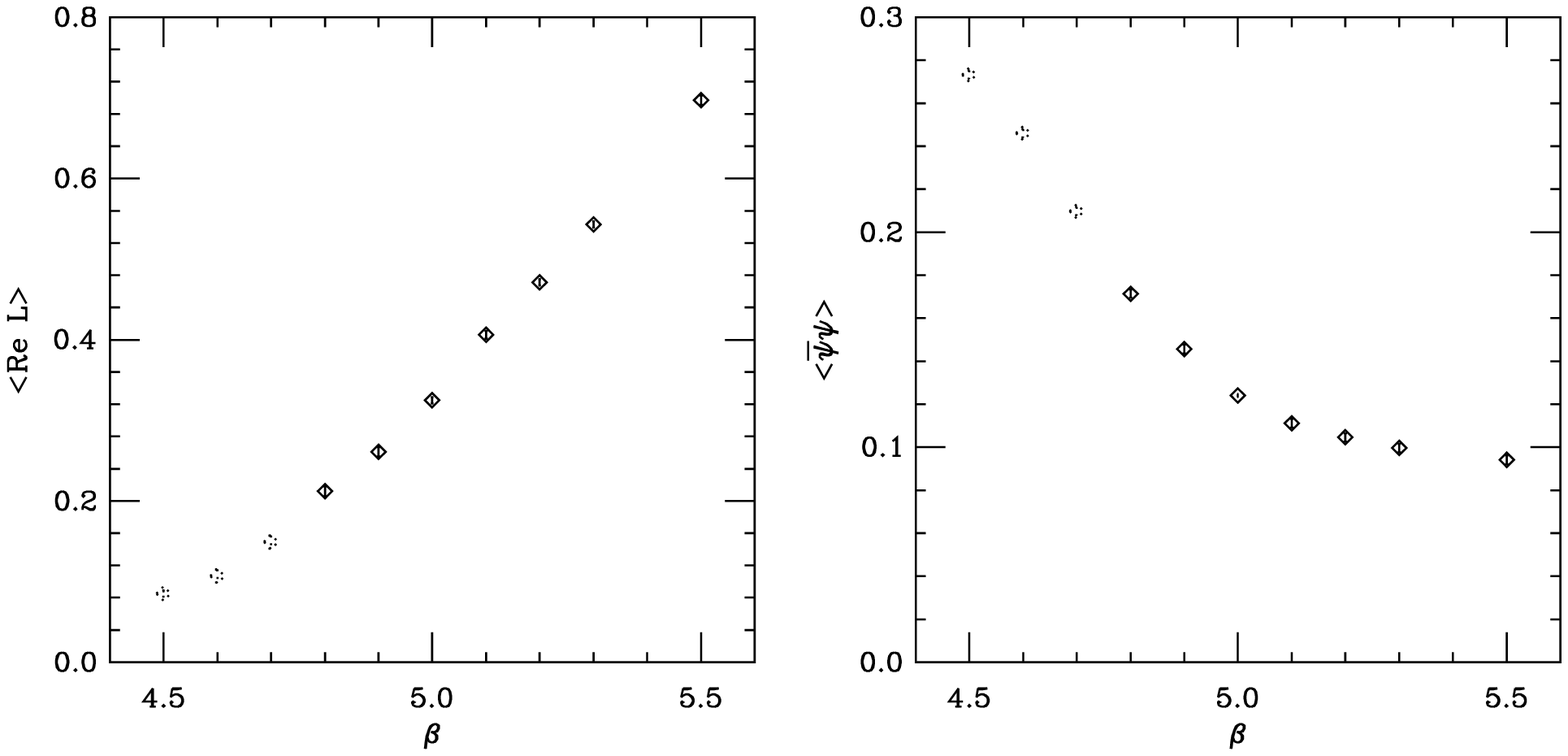}} \par}

\caption{HYP1 action simulated on \protect\( 8^{3}\times 4\protect \) lattices
at quark mass \protect\( am_{q}=0.1\protect \) (Set 1 in table \ref{mass_values}).
The real part of the Polyakov line and the chiral condensate \protect\( <\bar{\psi }\psi >\protect \)
are plotted as a function of \protect\( \beta \protect \). Dotted
symbols indicate data points where the lattice spacing is too large
to expect continuum like behavior.\label{HYP1_0.1}}
\end{figure}

\subsection{Phase structure and flavor symmetry}

\begin{table}
{\centering \begin{tabular}{|c||c|c||c|c|c||c|c|}
\hline 
Action&
\( \beta  \)&
\( am_{q} \)&
\( r_{0}/a \)&
\( \sqrt{\sigma }a \)&
\( a\,  \){[}fm{]}&
\( \Delta _{\pi } \)&
\( m_{G}/m_{\rho } \)\\
\hline
\hline 
HYP1&
5.0&
0.1&
-&
0.51(3)&
0.23(2)&
0.144(6)&
0.66(2)\\
\hline 
HYP1&
5.2&
0.1&
3.03(4)&
0.37(1)&
0.16&
0.057(6)&
0.719(8)\\
\hline
\hline 
thin&
5.2&
0.06&
2.85(2)&
0.40(1)&
0.18&
0.63(3)&
0.56(2)\\
\hline
\end{tabular}\par}

\caption{Physical scale and flavor symmetry violation of the dynamical HYP1
action. For comparison we list the corresponding values for the thin
link action whose bare parameters are chosen to approximately match
the physical scale and Goldstone pion mass of the \protect\( \beta =5.2\protect \),
\protect\( am_{q}=0.1\protect \) HYP1 simulations. \label{hypscale}}
\end{table}
Let us look first at the largest mass set, Set 1 in table \ref{mass_values}.
Figures \ref{standard}, \ref{APE1} and \ref{HYP1_0.1} show the
real part of the Polyakov line and \( <\bar{\psi }\psi > \) for the
three actions on \( 8^{3}\times 4 \) lattices. The thin link action
shows a pronounced discontinuity at \( \beta \simeq 5.1 \). The jump
in the Polyakov line is about 0.4, in \( <\bar{\psi }\psi > \) about
0.2. The APE1 action also gives indication of a first order phase
transition at \( \beta \simeq 4.85 \) but the discontinuity of the
Polyakov line is only about 0.2, of \( <\bar{\psi }\psi > \) about
0.1. Also, the discontinuity is at a lower \( \beta  \) value with
respect to the thin link action. Although we do not have a scale determination
for the APE1 action, it is unlikely that the physical value of the
phase transition temperature is unchanged. As figure \ref{HYP1_0.1}
shows the discontinuity completely disappears with the HYP1 action.
Both the deconfinement and chiral order parameters change smoothly
from the low to the high temperature phase. Moreover the susceptibility
\( \chi _{\bar{\psi }\psi ,\, disc} \) does not show a peak as a
function of the coupling \( \beta  \) in the investigated range.

The observation that the phase transition weakens as the gauge connections
between fermions become smoother and eventually disappears for our
smoothest HYP1 action is a strong indication that the phase transition
of the thin link action is a lattice artifact. To support this further
we determined the physical scale of the HYP1 action at two coupling
values, \( \beta =5.0 \) and \( \beta =5.2 \), with mass \( am_{q}=0.1 \)
on \( 8^{3}\times 24 \) lattices. On the same configurations we also
measured the meson spectrum. Our results are summarized in table \ref{hypscale}
where we list the values for the Sommer scale \( r_{0}/a \), the
string tension \( a\sqrt{\sigma } \), the relative mass splitting
\( \Delta _{\pi }=(m_{\pi }-m_{G})/m_{G} \) between the lightest
non-Goldstone pion \( \pi _{i,5} \) and the Goldstone pion \( G \)
and the pion to rho mass ratio. At \( \beta =5.0 \) the violations
of rotational symmetry in the static potential are substantial and
we were not able to determine reliably \( r_{0}/a \). We could obtain
a value for the string tension though from which we get the lattice
spacing quoted in table \ref{hypscale}. At \( \beta =5.2 \) both
quantities \( r_{0}/a \) and \( a\sqrt{\sigma } \) give a consistent
value for the lattice spacing. For comparison in table \ref{hypscale}
we also give the corresponding values for the thin link action at
\( \beta =5.2 \), \( am_{q}=0.06 \). This quark mass value corresponds
approximately to the \( am_{q}=0.1 \) HYP bare mass. As the results
in table \ref{hypscale} show the scale of the HYP and thin link actions
are fairly close at the same gauge coupling once the bare quark mass
values are matched. 

Again, we see that the flavor symmetry restoration predicted by quenched
simulations agrees with the dynamical results. The quenched study
of Ref. \cite{Hasenfratz:2001hp} predicted, at a lattice spacing
\( a\simeq 0.17\,  \)fm, a mass splitting \( \Delta _{\pi }=0.049(8) \)
with the HYP1 valence action at \( am_{q}=0.1 \). As table \ref{hypscale}
shows the dynamical simulations confirm this result. Flavor symmetry
with the HYP1 action is an order of magnitude better than with the
thin link action. 

The lattice spacing of the HYP action at \( \beta =5.0 \), \( am_{q}=0.1 \)
is already \( 0.23 \)fm, 1.4 times larger than the lattice spacing
at \( \beta =5.2 \), \( am_{q}=0.1 \). Naive scaling implies that
the lattice spacing is above \( a=0.3 \)fm at \( \beta =4.8 \).
Considering even lower \( \beta  \) values does not make much sense
physically. Once the lattice spacing is comparable or larger than
the typical instantons on the configuration one cannot expect to see
continuum like chiral behavior. We used dotted symbols in figure \ref{HYP1_0.1}
to plot the \( \beta <4.8 \) results to indicate that those data
points are not likely to show continuum behavior. 
\begin{figure}
{\centering \resizebox*{16cm}{!}{\includegraphics{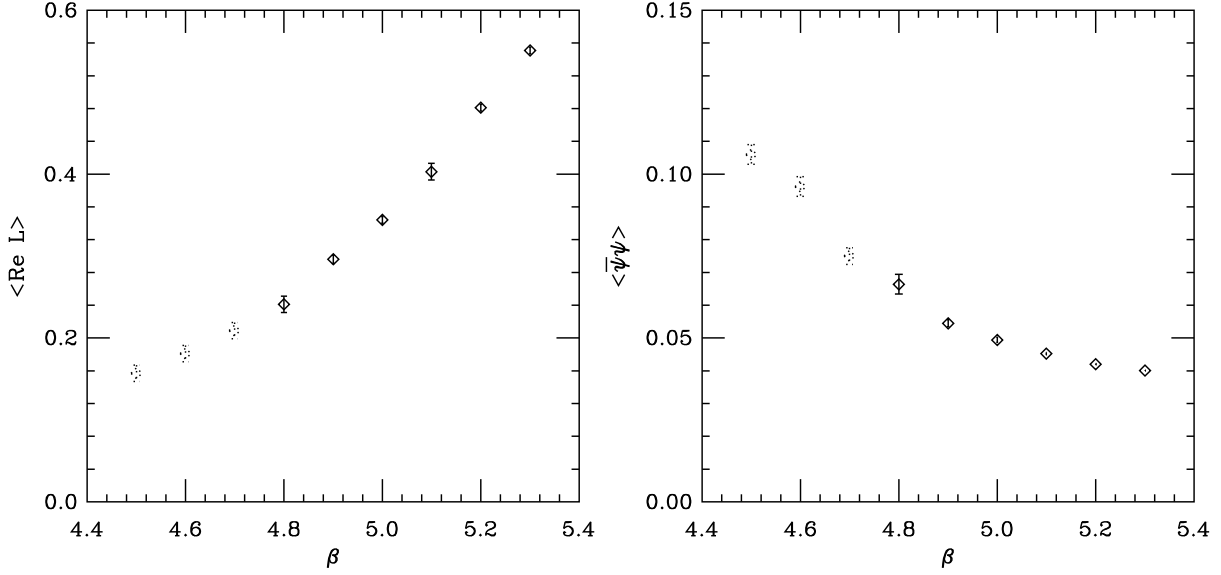}} \par}

\caption{HYP1 action simulated on \protect\( 8^{3}\times 4\protect \) lattices
at quark mass \protect\( am_{q}=0.04\protect \) (Set 2 in table \ref{mass_values}).
The real part of the Polyakov line and the chiral condensate \protect\( <\bar{\psi }\psi >\protect \)
are plotted as a function of \protect\( \beta \protect \). Dotted
symbols indicate data points where the lattice spacing is too large
to expect continuum like behavior.\label{HYP1_0.04}}
\end{figure}

\begin{figure}
{\centering \resizebox*{16cm}{!}{\includegraphics{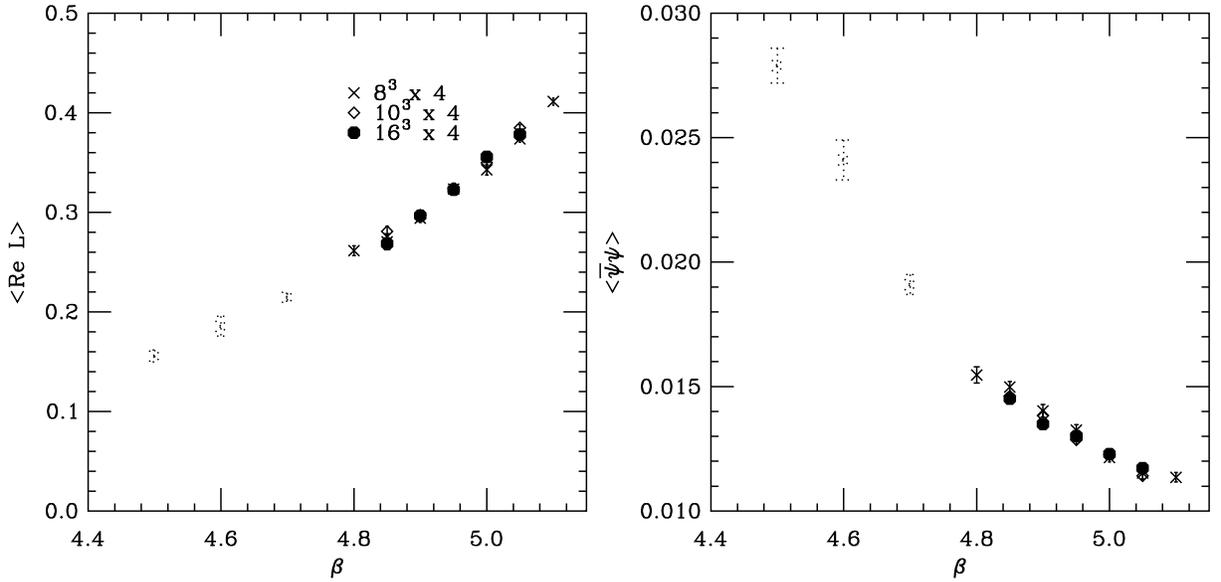}} \par}

\caption{HYP1 action simulated on \protect\( 8^{3}\times 4\protect \) (crosses),
\protect\( 10^{3}\times 4\protect \)(diamonds) and \protect\( 16^{3}\times 4\protect \)
(filled octagons) lattices at quark mass \protect\( am_{q}=0.01\protect \)
(Set 3 in table \ref{mass_values}). The real part of the Polyakov
line and the chiral condensate \protect\( <\bar{\psi }\psi >\protect \)
are plotted as a function of \protect\( \beta \protect \). Dotted
symbols indicate data points where the lattice spacing is too large
to expect continuum like behavior.\label{HYP1_largeV}}
\end{figure}

\subsection{The phase diagram with the HYP action}

The HYP1 action shows only a crossover with \( am_{q}=0.1 \) quark
mass. This might not be that surprising if we look at the endpoints
of the first order phase transition lines for the thin link action
quoted in \cite{Gottlieb:1991by}. At \( N_{t}=4 \), \( am_{max}\simeq 0.073 \)
while at \( N_{t}=6 \), \( am_{max}\simeq 0.021 \), showing a substantial
scale breaking. The physical mass of Set 1 is lower than \( m_{max} \)
at \( N_{t}=4 \) but larger than \( m_{max} \) one would predict
from the \( N_{t}=6 \) simulations assuming scaling. 

In this section we consider the HYP1 action at smaller quark masses
to see if we can find a first oder phase transition. The quark mass
of Set 2 is, in physical units, below the observed endpoint \( m_{max} \)
of the \( N_{t}=6 \) thin link transition. Numerical results for
the thin link action at \( am_{q}=0.024 \) on \( 8^{3}\times 4 \)
lattices show a strong first order phase transition. Yet the results
with the HYP1 action on \( 8^{3}\times 4 \) lattices plotted in figure
\ref{HYP1_0.04} show only a very broad crossover. 

Looking for the phase transition signal we simulated the HYP1 action
at the even lower quark mass of Set 3, \( am_{q}=0.01 \). At this
low value of the quark mass finite volume effects could be important.
For example the first order phase transition observed with the thin
link action at \( N_{t}=6 \) on \( 16^{3}\times 6 \) lattices at
a quark mass \( am_{q}=0.01 \) in Ref. \cite{Brown:1990by} washes
away if we repeat the simulations on \( 8^{3}\times 6 \) lattices.
To make sure that our results are not due to finite size effects we
simulated the HYP1 action at quark mass \( am_{q}=0.01 \) on \( 8^{3}\times 4 \),
\( 10^{3}\times 4 \) and \( 16^{3}\times 4 \) lattices. The results
are shown in figure \ref{HYP1_largeV}. There is no sign of significant
deviations as the spatial volume is increased by a factor of eight
and the phase diagram shows a very broad crossover. We conclude that
there is no signal for a first order phase transition at \( N_{t}=4 \)
for quark masses down to approximately the physical light quark mass
in the physically relevant gauge coupling range.

\begin{figure}
{\centering \resizebox*{10cm}{!}{\includegraphics{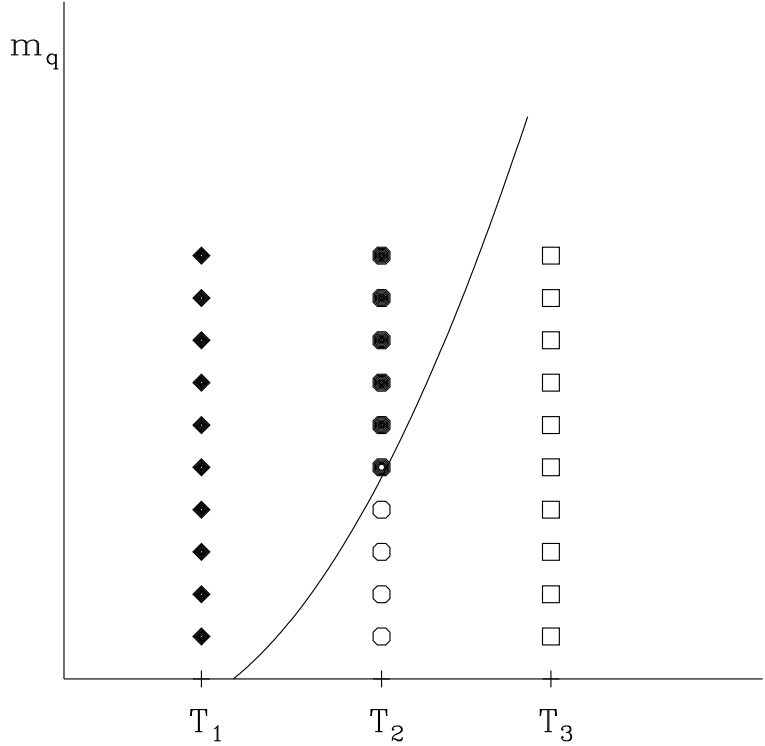}} \par}

\caption{A schematic phase diagram in the temperature - quark mass plane.
Open symbols are in the chirally symmetric, filled symbols are in
the chirally broken phase. The solid line represents the phase transition
or crossover between the phases.\label{phase_diagram}}
\end{figure}

Since we found no sign of a phase transition, only a gradual change
in the Polyakov loop and chiral condensate expectation values, one
might wonder if we found a chirally broken phase at all. To study
this question we consider the chiral condensate at fixed gauge coupling
as the function of the bare quark mass. For small quark mass the condensate
is expected to be linear in the mass suggesting to look at the quantity
\( <\bar{\psi }\psi >/m_{q} \) as the function of \( 1/m_{q} \)\begin{equation}
\label{lin_extrap}
\frac{<\bar{\psi }\psi >}{m_{q}}=\frac{\Sigma }{m_{q}}+c
\end{equation}
 where \( \Sigma  \) is the value of the condensate extrapolated
to zero quark mass. Figure \ref{phase_diagram} shows the possible
situations one might encounter. The solid line in the figure separates
the chirally broken (low temperature) and chirally symmetric (high
temperature) phases. This line could describe a phase transition or
only a crossover. The points at temperature \( T_{3} \) are in the
chirally symmetric phase, \( \Sigma =<\bar{\psi }\psi >|_{m_{q}=0}=0 \),
the linear form of eq.(\ref{lin_extrap}) should predict a constant
value for \( <\bar{\psi }\psi >/m_{q} \). The points at \( T_{1} \)
are all in the chirally broken phase, \( \Sigma  \) is finite and
positive, eq.(\ref{lin_extrap}) should predict a linear dependence
for \( <\bar{\psi }\psi >/m_{q} \) with a positive slope. The case
at \( T_{2} \) is more complicated. The points below the solid line
(open symbols) are in the chirally symmetric phase and should behave
the same as the open symbol points at \( T_{3} \) giving \( \Sigma =0 \)
in the extrapolation. The points above the solid line (filled symbols)
are in the chirally broken phase and the linear form eq.(\ref{lin_extrap})
should describe the mass dependence in the chirally broken phase with
a positive condensate at the transition point. The quantity \( \Sigma  \)
in eq.(\ref{lin_extrap}) however is not this quantity but its linearly
extrapolated value to zero quark mass. \( \Sigma  \) could therefore
take any value, even negative ones, if at the transition point the
condensate is small or the quark mass is large. 
\begin{figure}
{\centering \resizebox*{10cm}{!}{\includegraphics{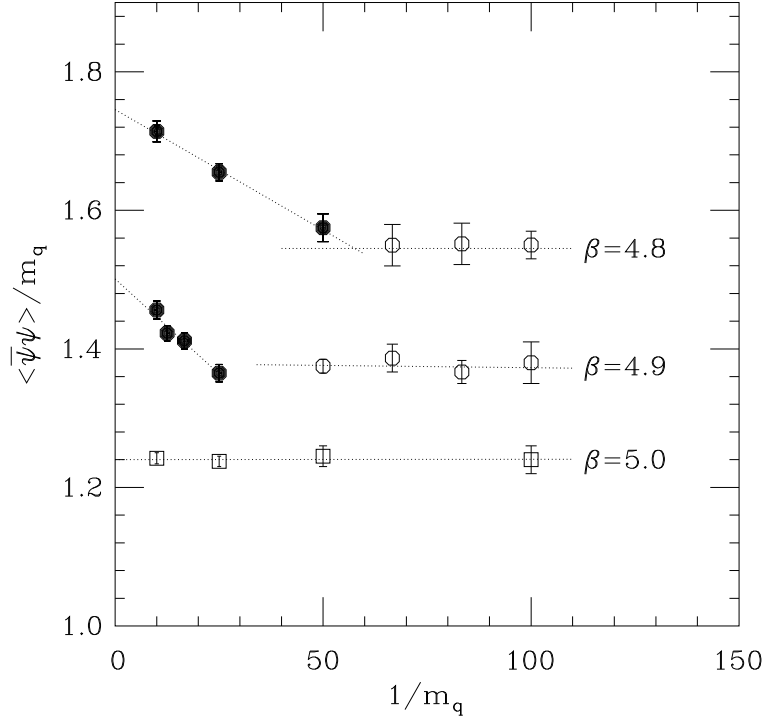}} \par}

\caption{\protect\( <\bar{\psi }\psi >/m_{q}\protect \) as the function of
\protect\( 1/m_{q}\protect \) at \protect\( \beta =4.8\protect \),
\protect\( \beta =4.9\protect \) and \protect\( \beta =5.0\protect \).
All data are on \protect\( 8^{3}4\protect \) lattices.\label{pbp_extrapolation}}
\end{figure}

We have analyzed the chiral condensate on the \( 8^{3}\times 4 \)
configurations according to eq.(\ref{lin_extrap}). In addition to
the quark mass values we presented above we simulated the system at
a few additional ones to make the extrapolation meaningful. Figure
\ref{pbp_extrapolation} shows \( <\bar{\psi }\psi >/am_{q} \) as
the function of \( 1/am_{q} \) at \( \beta =4.8 \) , \( \beta =4.9 \)
and \( \beta =5.0 \). These \( \beta  \) values correspond to the
situations described at \( T_{2}, \) \( T_{2} \) and \( T_{1} \),
respectively, in figure \ref{phase_diagram}. For easier identification
we have used the same symbols in the two figures. At \( \beta =4.8 \)
and 4.9 the slope \( \Sigma  \) is negative for \( am_{q}\geq 0.025(0.03) \),
and \( \Sigma  \) is consistent with zero for \( am_{q}<0.025(0.03) \)
indicating that there is a phase boundary separating the chirally
broken and symmetric phases around \( am_{q}\approx 0.025(0.03) \).
We have found negative \( \Sigma  \) slope at all \( \beta \leq 4.9 \)
coupling though the errors at \( \beta \leq 4.7 \) are too large
to distinguish \( \Sigma  \) from zero. At \( \beta =5.0 \) all
data points are consistent with a vanishing slope, \( \Sigma =0 \),
suggesting that all points are in the chirally symmetric phase. We
have found similar behavior for \( \beta \geq 5.0 \). These observations
suggest that the simulations have found the chirally broken phase
with all investigated quark mass values. They also suggest that the
chiral condensate at the crossover is small and that the chiral \( am_{q}=0 \)
phase transition, if exists, is at a fairly low \( \beta  \) value
on \( N_{t}=4 \) lattices. These results are consistent with the
interacting instanton liquid model predictions \cite{Schafer:1996pz, Schafer:1998wv}
but not with the thin link action lattice results.
\begin{figure}
{\centering \resizebox*{16cm}{!}{\includegraphics{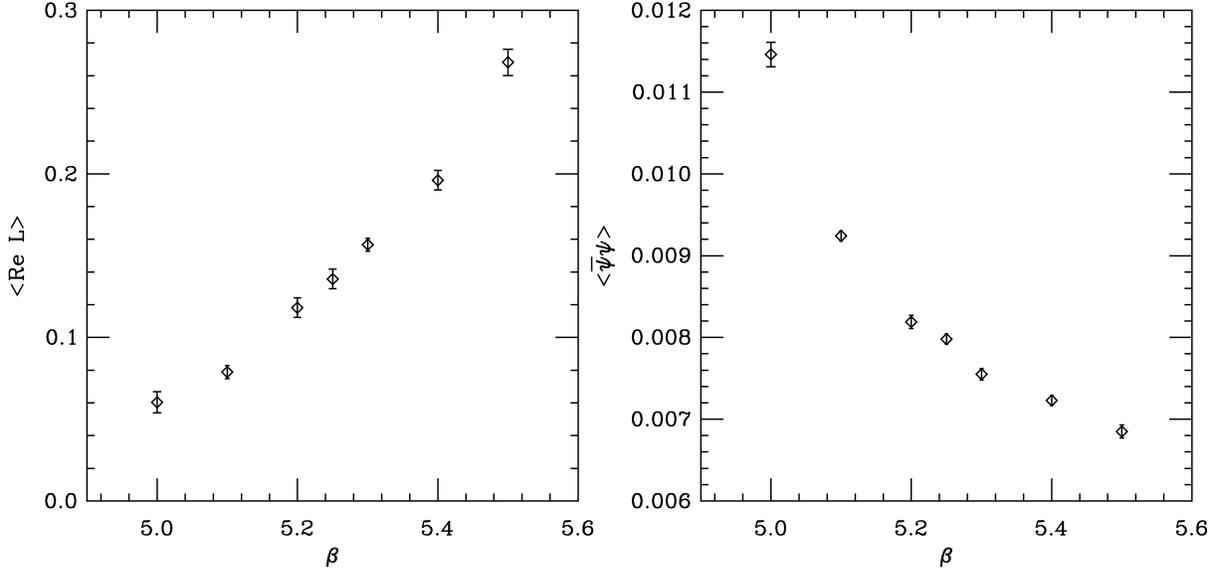}} \par}

\caption{HYP1 action simulated on \protect\( 16^{3}\times 6\protect \) lattices
at quark mass \protect\( am_{q}=0.0067\protect \) (Set 3 in table
\ref{mass_values}). The real part of the Polyakov line and the chiral
condensate \protect\( <\bar{\psi }\psi >\protect \) are plotted as
a function of \protect\( \beta \protect \).\label{nt6thermo}}
\end{figure}

In order to understand the phase structure of four flavor fermions
it would be important to determine the temperature at the \( am_{q}=0 \)
phase transition point and locate the phase transition line at finite
quark mass. This, however, should not be done on \( N_{t}=4 \) lattices.
As we pointed out in the introduction, simulations at \( N_{t}=4 \)
might be questionable even with chirally symmetric actions because
the lattice spacing is too large to support the correct topological
structure of the vacuum. Furthermore, decreasing the quark mass while
staying at \( N_{t}=4 \) makes flavor symmetry violations larger
even with the HYP1 action. We believe that the disappearance of the
first order phase transition is due to the improved flavor symmetry
of the HYP1 action. These considerations indicate that we have to
consider lattices with \( N_{t}\geq 6 \) temporal extension.

In figure \ref{nt6thermo} we show our first results for \( N_{t}=6 \)
thermodynamics with HYP1 action on \( 16^{3}\times 6 \) lattices
at a quark mass \( am_{q}=0.0067 \). This mass matches approximately
the lightest quark mass of Set 3 at \( N_{t}=4 \). The critical region,
which we identify as the region where \( <\bar{\psi }\psi > \) starts
increasing and where the autocorrelation times are the largest, moves
to higher \( \beta  \) values than at \( N_{t}=4 \). Again, there
is no sign of a first order phase transition, the Polyakov line and
the chiral condensate are smooth, indicating a crossover as observed
at \( N_{t}=4 \). The susceptibility \( \chi _{\bar{\psi }\psi ,\, disc} \)
does not show any peak in the investigated range. With a shift \( \Delta \beta \approx 0.2 \)
the susceptibility agrees with the one obtained on the \( N_{t}=4 \)
lattices. Also the chiral condensate normalized by the quark mass,
\( <\bar{\psi }\psi >/m_{q} \) matches the \( N_{t}=4 \), \( am_{q}=0.01 \)
data if the gauge coupling is shifted by \( \Delta \beta =0.2 \).
This shift in the gauge coupling for a scale change of \( \frac{6}{4}=1.5 \)
is consistent with what we observed between the \( \beta =5.0 \)
and \( \beta =5.2 \), \( am_{q}=0.1 \) zero temperature simulations. 

Our \( N_{t}=6 \) results are only preliminary. Obviously the simulations
should be repeated at different quark mass values in order to do the
extrapolation according to eq. (\ref{lin_extrap}) but that is beyond
the scope of this paper. Our goal here was to demonstrate the striking
difference between the thin link and HYP actions.

\section{Summary}

In this paper we studied the finite temperature phase diagram of four
flavor fat link staggered fermion actions on lattices with \( N_{t}=4 \)
and \( N_{t}=6 \) temporal extents. We used a new smearing, the hypercubic
blocking (HYP), for the fat links. Hypercubic blocking mixes the gauge
links only within hypercubes attached to the original link minimizing
thereby the lattice artifacts arising from extended smearing. The
action with one level of hypercubic blocking (HYP1) considerably reduces
the flavor symmetry breaking of the standard thin link action but
does not change the local properties of the gauge fields beyond distances
\( r/a\geq 2 \). We presented a new algorithm for simulating the
HYP1 action. This algorithm is very simple because it is based on
the overrelaxation and heatbath algorithms for the pure gauge action.
It is efficient as the quark mass is lowered allowing simulations
which would be impractical with thin link action. 

We found that improved flavor symmetry, presumably the presence of
15 near-degenerate Goldstone bosons, changes the finite temperature
phase diagram of four flavor QCD. Our simulations with the HYP1 action
do not show any sign of a first order phase transition even at physical
quark masses of a few MeV. We found indications for a broad crossover-like
transition between the chirally broken and symmetric phases. The phase
transition temperature at zero quark mass is too small and cannot
be identified on \( N_{t}=4 \)  lattices even with chirally improved
actions.

Our algorithm to simulate the HYP1 action can be generalized to arbitrary
number of flavors by using a polynomial approximation for the fermionic
action \cite{colonext}.

\section{Acknowledgements}

We are indebted to M. Hasenbusch for discussions that lead to the
new algorithm presented here. We benefited from many discussions with
Prof. T. DeGrand during the course of this work. The simulations were
performed in scalar and parallel mode on the beowulf cluster of the
high energy theory group and on the linux farm of the high energy
experimental group at University of Colorado. We would like to thank
the system administrator, D. Johnson, for his precious technical support.
This work was supported by the U. S. Department of Energy. Finally
we thank the MILC collaboration for the use of their computer code.

\bibliographystyle{apsrev}
\bibliography{lattice}

\end{document}